# A Systematic Review of the Efforts and Hindrances of Modeling and Simulation of CAR T-cell Therapy

**Ujwani Nukala[1], Marisabel Rodriguez Messan[1], Osman N. Yogurtcu[1], Xiaofei Wang[2], Hong Yang[1*]**

**1 Office of Biostatistics and Epidemiology, Center for Biologics Evaluation and Research, US FDA, Silver Spring, MD, USA**
**2 Office of Tissues and Advanced Therapies, Center for Biologics Evaluation and Research, US FDA, Silver Spring, MD, USA**

Chimeric Antigen Receptor (CAR) T-cell therapy is an immunotherapy that has recently become highly instrumental in the fight against life-threatening diseases. A variety of modeling and computational simulation efforts have addressed different aspects of CAR T-cell therapy, including T-cell activation, T- and malignant cell population dynamics, therapeutic cost-effectiveness strategies, and patient survival. In this article, we present a systematic review of those efforts, including mathematical, statistical, and stochastic models employing a wide range of algorithms, from differential equations to machine learning. To the best of our knowledge, this is the first review of all such models studying CAR T-cell therapy. In this review, we provide a detailed summary of the strengths, limitations, methodology, data used, and data gap in current published models. This information may help in designing and building better models for enhanced prediction and assessment of the benefit-risk balance associated with novel CAR T-cell therapies, as well as with the data need for building such models.

**Keywords:** CAR T-cell therapy, Modeling and simulation, Survival analysis, Cost-effectiveness, Pharmacokinetics-pharmacodynamics.

## INTRODUCTION

Despite recent therapeutic advances in treating cancer, issues such as variable treatment responses, high rate of relapse, and poor prognosis in relapsed/refractory (r/r) cancers continue to be major challenges. The Chimeric Antigen Receptor (CAR) T-cell therapy is a promising form of adoptive immune cell therapy. In this therapy, patient T-cells are genetically modified to express CAR, consisting of a surface antigen-targeting antibody single chain variable fragment (scFv) fused to a signaling domain of the T-cell receptor (TCR). These genetically-modified T-cells target and eliminate cancer cells expressing specific antigens. Since its first *in vitro* proof of concept demonstration on mouse MD45 cells in 1989 (1), CAR T-cell therapy has evolved significantly and has recently become available for medical use for a subset of cancer types and patients. In 2017, via Breakthrough Therapy Designation, the first autologous CAR T-cell therapies, tisagenlecleucel (tisa-cel) (KYMRIAH) (2) and axicabtagene ciloleucel (axi-cel) (YESCARTA) (3) for treating patients with hematological malignancies were approved, followed by the recent approval of brexucabtagene autoleucel (TECARTUS) in 2020 and all were enthusiastically received. The increasing interest in CAR T-cell therapy is evident from the number of clinical trials; more than 700 registered CAR T therapy-based clinical trials were listed on ClinicalTrials.gov by the end of October 2020. CAR T-cells targeting different cancer types, including solid tumors of the liver, brain, breast (e.g. ClinicalTrials.gov

identifier: NCT02541370), lung, pancreas (such as NCT02349724, NCT01869166), and many others (4), as well as infectious diseases (5), are in developmental and evaluation stages.

## Modeling and its Importance for CAR T-Cell Therapy

Mathematical models have been widely used in biological and medical research. These models generate or validate hypotheses from experimental data, predict different possible outcomes through *in silico* simulations and are among the most prominent methods used to study several aspects of diseases, such as the interplay of complex immune responses and drugs/biologics (6, 7). Different modeling approaches can be considered to explore issues in drug development**,** including but not limited to disease progression (8-12), immune response to immunotherapy (13-17), and others (18-22). FDA's recent Model-Informed Drug Development Pilot Program facilitates the development and application of exposure-based, biological, and statistical models derived from preclinical and clinical data sources. It also uses a variety of quantitative methods to help balance the risks and benefits of drug products in development (23).

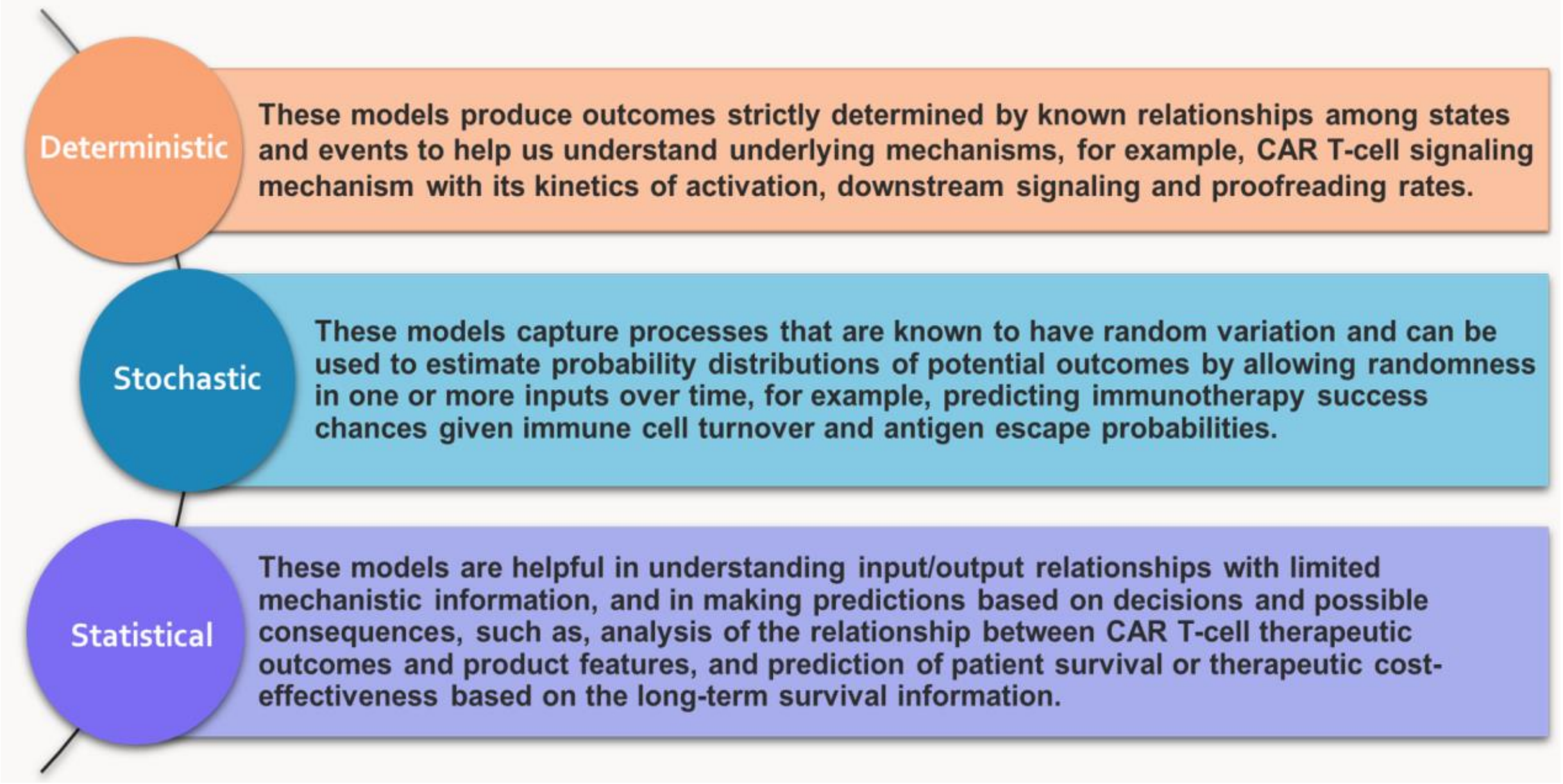

**Figure 1. Major modeling frameworks addressing different aspects of CAR T-cell therapy**

Cancer biology involves complex dynamic interactions of cancer cells with their surrounding tissue microenvironment. Understanding the dynamics and interactions between the tumor cells, and the human immune system provides insight into how specific interventions such as immunotherapy aid in fighting cancer (24). Some deterministic, stochastic and statistical models have been published to analyze the cellular kinetics, cost-effectiveness, and patient survival statistics of CAR T-cell therapy specifically. These models have become instrumental in understanding the complex dynamics and interactions involved in CAR T-cell therapy (Figure 1). Cancer biology involves complex dynamic interactions of cancer cells with their surrounding tissue microenvironment. Understanding the dynamics and interactions between the tumor cells, and the human immune system provides insight into how specific interventions such as immunotherapy aid in fighting cancer (24). Some deterministic, stochastic and statistical models have been published to analyze the cellular kinetics, cost-effectiveness, and patient survival statistics of CAR T-cell therapy specifically. These models have become instrumental in understanding the complex dynamics and interactions involved in CAR T-cell therapy (Figure 1). There have been reviews highlighting, for instance, the use of mathematical models in cancer (25) and demonstrating the utility of ordinary differential equation-based mechanistic models to study T-cell activation (26). Notably, Markaryan et al. (25) provided a very detailed description of proposed models covering single cells, multicellular interactions, and multiscale/spatio-temporal frameworks to better understand the complex tumor microenvironment and Chaudhury et al. (27) presented a review of cellular kinetic-pharmacodynamic modeling approaches. However, no comprehensive reviews specifically focus on various modeling efforts for CAR T-cell therapy. This review of computational models studying various aspects of CAR T-cell therapy could help modelers better comprehend available modeling approaches, relevant research questions, and existing knowledge gaps. Additionally, the current review may inform clinical researchers about the importance of collectable clinical trial data for development of better models for benefit-risk assessment of CAR T-cell therapy.

## LITERATURE SEARCH METHODOLOGY

Using a customized search query developed by an expert FDA librarian, we conducted a literature search of five scientific publication databases (PubMed, Embase, Web of Science, Google Scholar and BioRxiv) limiting the search to English language and identified 360 articles. The authors UN, MRM, and ONY after removing duplicates and carefully reviewing all article titles and abstracts, selected 26 articles for this in-depth review. These 26 articles were selected as they were studying different aspects of CAR T-cell therapy (such as therapeutic efficacy, safety, and cellular kinetics) using various types of modeling and simulation approaches (found using keywords such as probabilistic, mathematical, statistical, and predictive). The authors UN, MRM, ONY, and XW independently prepared article summaries which were later discussed, compared and condensed. Purely exploratory noncompartmental analyses (NCA) and models that focused on molecular dynamics (MD) simulations of CAR molecules were considered beyond the scope of this review and excluded. Details of our custom search query and systematic literature search can be found in the supplementary text along with expansions of all acronyms used in this article.

## CAR T-CELL MODELING AND SIMULATION EFFORTS

In this section, we provide a general overview of modeling and simulation approaches assessing different dimensions of CAR T-cell therapy. Each article reviewed in this paper is unique in terms of its model/algorithm type, the data (at population, individual patient, or cellular levels), and modeling tools. In Table 1, we list the articles that were reviewed in detail, and in Figure 2 we provide a visual summary of various modeling and simulation efforts studying different aspects of CAR T-cell therapy.

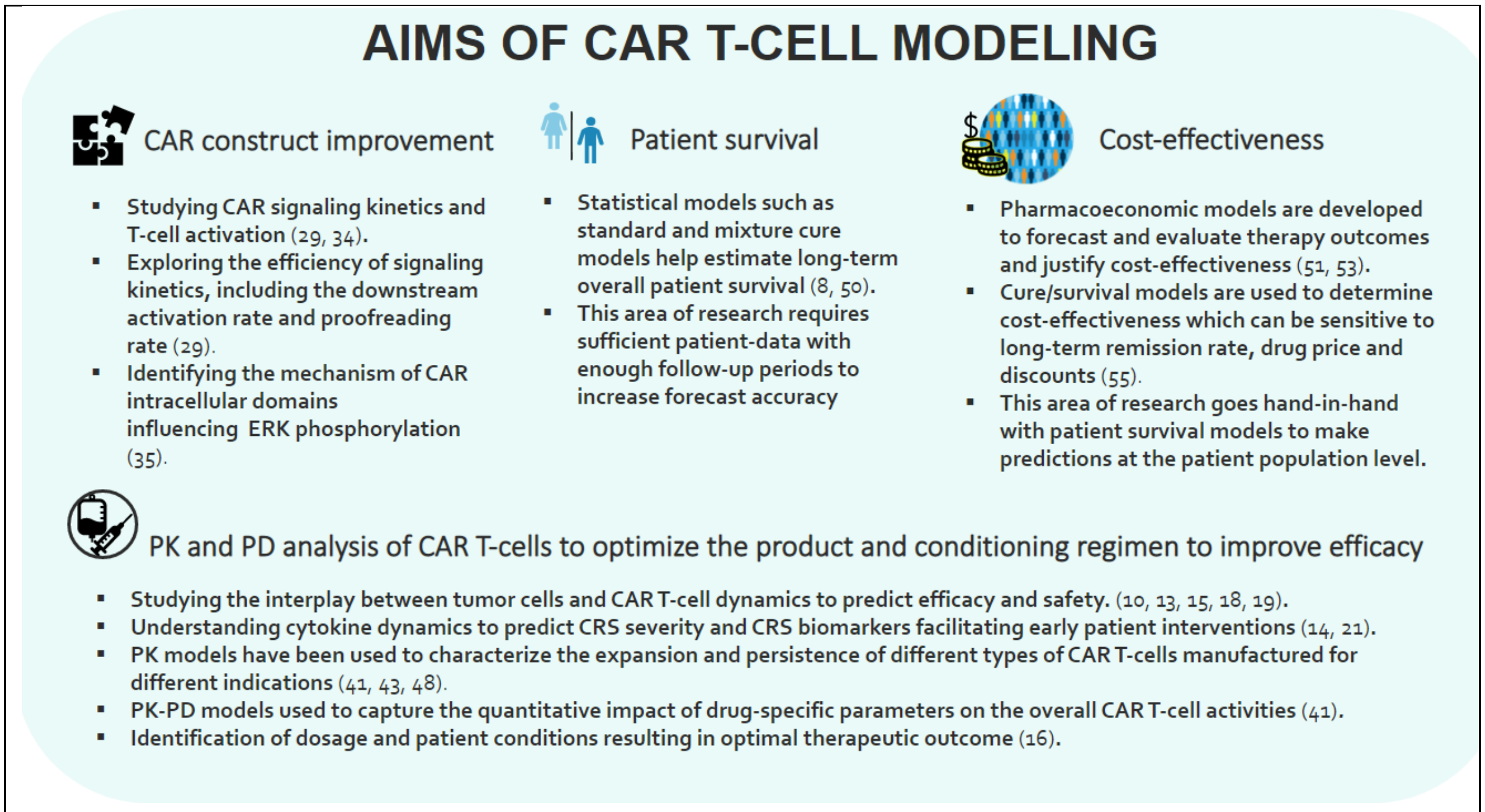

**Figure 2. CAR T-cell therapy modeling and simulation efforts covered in this review**

## CAR Molecules and Models of CAR T-Cell Activation Signaling Cascade

A CAR molecule consists of an intracellular signaling domain and an extracellular antigen-recognizing domain (ARD). As described in Figure 3, the extracellular domain carries a single chain variable fragment (scFv) antibody that is designed to bind to the target antigen. The CAR intracellular domain consists of a signaling domain, CD3ζ, which is derived from human TCR. This is the structure of the first generation of CARs. The next generations of CARs are equipped with a co-stimulatory domain, either CD28 or 4-1BB (second generation) or both CD28 and 4-1BB (third generation), in addition to the CD3ζ signaling domain, providing them with better proliferative capacity, tumoricidal activity, and increased cytokine secretion (28). The fourth generation CARs are additionally equipped with inducible expression cassettes for a transgenic protein (e.g., a cytokine). CAR T-cells can be directly activated via CAR in human leukocyte antigen (HLA)-independent manner, unlike TCR-redirected T-cells. Studying the underlying mechanisms of activation of CAR T-cells compared to natural response of TCR in the human immune system may be instrumental in optimizing CAR design and to extend the CAR T-cell therapies beyond hematological malignancies.

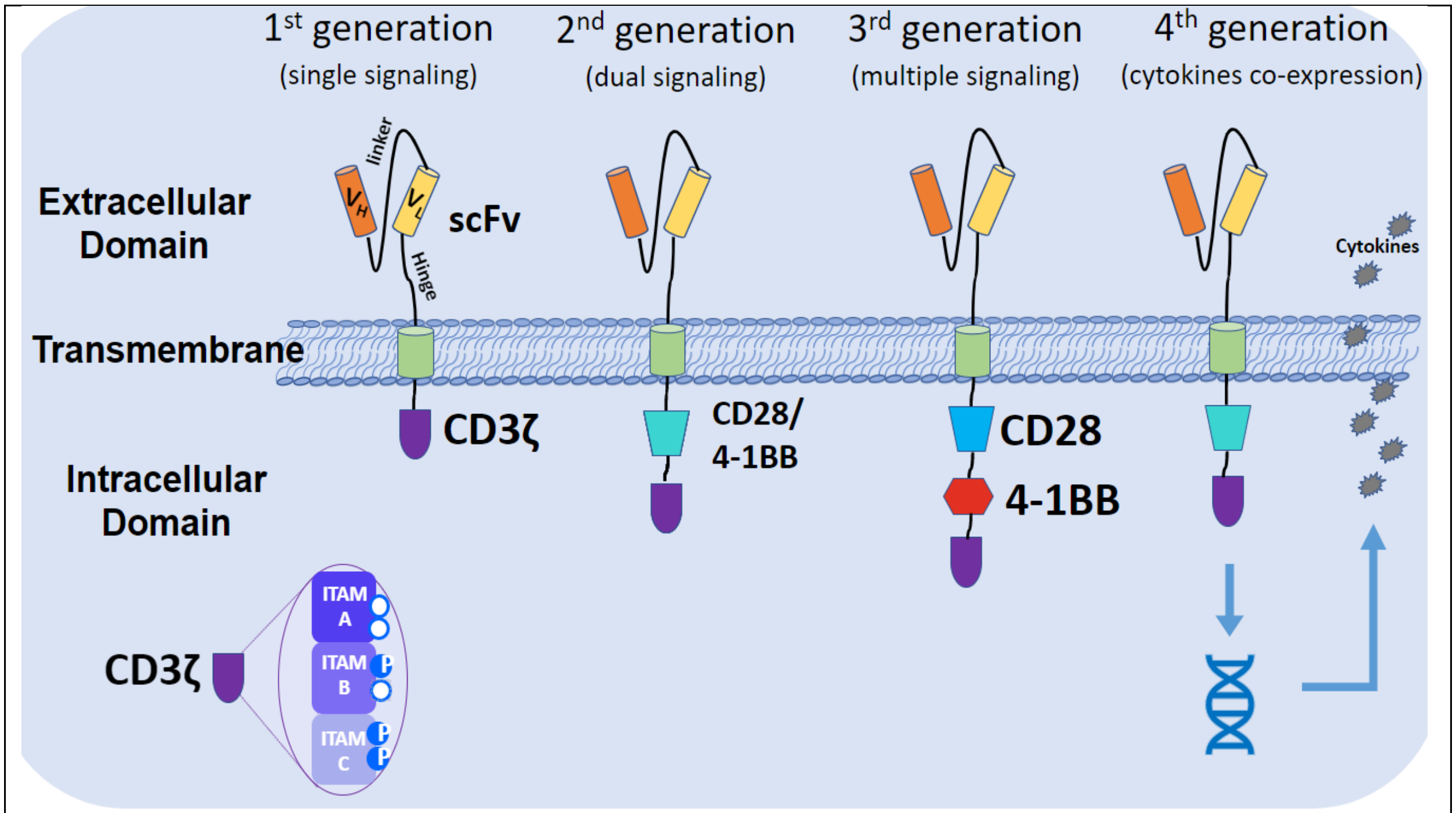

**Figure 3. Schematic diagram of different generations of CAR T-cell structure.** First generation CARs possess single intracellular signaling component, CD3ζ; Second generation CARs possess co-stimulatory domain, either CD28 or 41BB with CD3ζ; Third generation CARs possess additional co-stimulatory domain, CD28 and 41BB with CD3ζ; Fourth generation CARs inducibly express pro-inflammatory cytokines. CD3ζ domain has six tyrosine phosphorylation sites arranged in pairs on three immunoreceptor tyrosine-based activation motifs (ITAMs) and CD28 domain has two pairs of tyrosine phosphorylation sites; these sites are phosphorylated by LCK and the activated CD28 and CD3ζ domains further activate intermediate pathways which leads to ERK activation (arrows indicate activating interactions).

Harris et al. (29) employed ODE-based mathematical modeling to better understand and contrast the sensitivity and signaling capacity of T-cells with CARs and T-cells with TCRs in mouse cells. The research team found that although CAR cell membrane surface expression was 10-fold higher than that of ordinary TCRs, they were 10–100-fold less sensitive than TCRs. Mathematical modeling demonstrated that lower CAR sensitivity could be attributed to less efficient signaling kinetics, including reduced kinetic proofreading rate, reduced activation rate, or a combination of both mechanisms. Furthermore, reduced cytokine secretion observed at high antigen density for both TCRs and CARs suggested a role for negative regulators in both systems. Interestingly, at high antigen density, CARs compared to TCRs lead to secretion of 1.5 to 2-fold higher levels of IL-6. The use of modeling alongside *in vitro* experiments makes this article compelling and its findings help CAR design optimization for better signaling kinetics. However, we do not know how this model predictions on T-cell binding to therapeutically relevant antigen targets (e.g. CD19) in mouse cells could be extrapolated to humans.

The CAR containing the signaling domain CD3ζ has six tyrosine phosphorylation sites arranged in pairs on three immunoreceptor tyrosine-based activation motifs (ITAMs) (Figure 3), which when doubly phosphorylated, become activated, bind to the signaling proteins and perpetuate downstream signaling.

Several modeling studies have attempted to better understand ITAM activation and subsequent downstream signaling (30-33). However, these studies did not account for the competitive factors that may influence site-specific phosphorylation and are not validated with site-specific phosphorylation data. Axi-cel, a second-generation CAR T contains CD28 as a co-stimulatory domain in addition to CD3ζ signaling domain. The CD28 domain contains four additional tyrosine sites which are phosphorylated by LCK. The mechanism of recruitment and competition for LCK by CD28 can alter the phosphorylation of CD3ζ affecting the downstream signaling. This downstream signaling controls T-cell activation responses including cytotoxicity, cytokine production, proliferation and persistence.

To understand all these better and to overcome the limitations of previous studies, Rohrs et al. (34) quantitatively measured the site-specific phosphorylation of CAR proteins by LCK over time, using phospho-proteomic mass spectrometry. They also explored the influence of individual tyrosine sites over the phosphorylation kinetics of other sites, by individually mutating each tyrosine site and then measuring the percent phosphorylation of the other sites over time. By fitting this experimental data to their four candidate ODE-based mechanistic models, Rohrs and colleagues generated and tested new hypotheses regarding the mechanisms by which LCK phosphorylates CD3ζ ITAMs. They identified that the competitive inhibition mechanism-based model, where the phosphorylated and unphosphorylated tyrosine sites interact while competing with each other, accurately fitted their experimental data and the data from the literature (33) with lowest error and lowest Akaike information criterion (AIC).

Authors also investigated the influence of the co-stimulatory domain CD28 on CAR phosphorylation using phosphor-proteomic mass spectrometry to quantify the site-specific phosphorylation levels of 28ζ and found that CD28 increases the phosphorylation rate of CD3ζ, and their model indicated that the increasing effective catalytic rate of LCK is the likely mechanism. Phenomenologically, the best representative model, competitive inhibition model, was selected out of three other candidate models, sequential-order model, random-order model, and phosphate-priming model, developed based on Michaelis-Menten kinetics aiming to explain the tyrosine phosphorylation rates of the CD3ζ-CD28 construct. These candidate models are publicly available. The strengths of this study are that authors quantified phosphorylation at individual tyrosine sites on the intact intracellular domain of the CAR protein and the mechanisms identified using their model may help engineer quicker phosphorylating CARs, for optimal activation of T-cells.

However, this integrative experimental and modeling approach (34) was not able to explain how the increase in CD3ζ phosphorylation affects downstream signaling of T-cell activation. In another article (35), the same team constructed a larger mechanistic mathematical model of a CAR T-cell activation signaling cascade, including signaling initiated by antigen binding to the CAR and culminated in ERK phosphorylation. This model helped explain how the CAR intracellular domains influence activation of MAPK signaling pathway, which leads to ERK phosphorylation, the pathway that mediates T-cell activation and proliferation. This is one of the first, detailed (23 proteins), experimentally-parametrized, and validated mechanistic models of the CAR T-cell intracellular signaling network. By combining different models (in other words an ensemble model) from the literature which study the various signaling networks, the authors predicted that CD28 primarily influences ERK activation by way of recruitment of LCK, which increases the kinetics of CD3ζ activation. Since these models are based only on part of the intracellular signaling network, we do not know the extent of the impact of other relevant proteins not included in the model that may have on T-cell activation. Also, from their ensemble model, it is not immediately clear how having alternative CAR constructs would alter the cellular activation. The model was not validated with *in vivo* data, which makes knowing their predictive power in real-world patient treatment scenarios difficult.

Later, using their mechanistic model of CAR T-cell signaling (35), Cess and Finley demonstrated the influence of cell-to-cell heterogeneity in protein expression levels on the activation of CAR T-cells (36). The team applied partial least-squares (PLS) technique to encode the Monte Carlo simulation results of previously-developed nonlinear ODE-based mechanistic model of CAR T-cell signaling (35). Their simulation pipeline uses a set of initial concentrations of 23 proteins, allowing for unphosphorylated proteins, e.g., extracellular-signal-regulated kinase (ERK), proteins with varying levels of phosphorylation, free proteins, and various protein complex species. The researchers were able to find that only the expressions of proteins relating directly to the receptor and the mitogen-activated protein kinase (MAPK) cascade, the upstream and downstream of the reaction network, respectively, are relevant to a T-cell's response. Additionally, the research team found that increasing the number of available receptors can inhibit a cell's ability to respond due to increasing strength of negative feedback from phosphatases. These predictions can be validated against *ex vivo* data and can be used to help select and develop the "best" CAR T-cells for treatment. On the other hand, the type of modeling in (36) might have limitations such as overfitting due to high number of parameters, and the PLS technique assumes linear relationships between inputs and outputs. However, authors claim that there was a reasonable approximation of linear relationship because of their PLS model's high accuracy, although collinearity tests are typically necessary to demonstrate and ascertain such a result.

To sum up, the studies reviewed in this section (29, 34-36) employed mathematical models to better understand the downstream signaling mechanisms of CAR T-cell activation. These studies identified the competitive inhibition mechanism of CAR protein phosphorylation, the influence of co-stimulatory domain on overall CAR protein phosphorylation and how the cell-to-cell heterogeneity in terms of protein expression levels can influence the CAR T-cell activation response. These findings facilitate a deeper understanding of the underlying mechanisms of cellular activity and pave the way for designing a CAR that can achieve optimal CAR T-cell activation.

## Models used to evaluate and predict CAR T-cell therapeutic safety and/or efficacy

CAR T-cell therapy has yielded unprecedented efficacy with sustained remissions in patients with refractory cancers. Two-year follow-up data from the international single-arm, multicenter ZUMA-1 clinical trial showed 58% (n = 59/101) of patients had a complete response, suggesting that axi-cel can induce a median overall survival of more than two years (39). A phase-2 clinical trial of tisa-cel in pediatric patients showed an overall response rate of 81.3% (n = 61/75) (37). In addition, a meta-analysis of 38 published clinical studies including 665 patients treated with CAR T-cells for ALL, chronic lymphocytic leukemia (CLL), and B cell lymphomas showed an overall pooled response rate (RR) of 72%. The response rate for ALL, CLL, and lymphoma was 81%, 70% and 68%, respectively (38).

Despite these promising results, the regulatory assessment of CAR T-cell therapies remains challenging since CAR T-cell therapy is a living drug with outcomes that are variable and highly patient-specific.

Although, CAR T-cell therapy has yielded unprecedented efficacy with sustained remissions in cancer patients with refractory diseases, there are several factors which may impede CAR T-cell therapy efficacy or increase the risk of life-threatening immune-mediated adverse events. Utilizing mathematical predictive models, it may be possible to identify the factors that govern non-uniform CAR T-cell therapy efficacy and safety. Also, those models may be able to identify patients who are more likely to have successful therapeutic outcomes with tolerable toxicities, or those who may be at risk of relapse or severe Cytokine Release Syndrome (sCRS), so that risk mitigation strategies can be planned, possibly reducing the complications related to CRS symptoms and mortality rate. Though CAR T-cell therapy has demonstrated remarkable antitumor efficacy in treating B cell malignancies, its use in treating solid tumors remains in early stages of clinical development. The variety of CARs for treating different types of solid tumors, different target patient populations, and different preconditioning regimens make determining critical factors for CAR T-cell efficacy challenging (20). In this section, we summarized mathematical or pharmacokinetic models (using data either from animal models or clinical studies) focused on studying the interplay between tumor cells and CAR T-cells and/or the cytokines to help predict the efficacy and/or safety outcomes of CAR T-cell therapy.

Using classification and regression tree (CART) clustering, Finney et al. (9) analyzed the relationship between therapeutic outcomes (such as duration of leukemia-free survival and B cell aplasia) and product features (e.g., phenotype, function, and expansion of CAR T-cell). Their analysis showed that treatment outcome may be predictable (with $r^2=0.636$) based on expression levels of LAG-3 and TNF-α biomarkers in $CD8^+$ T-cells in the apheresis starting material, thus facilitating patient selection and clinical management accordingly.

George and Levine (10) developed the first stochastic modeling framework of co-evolution of the cancer cells, along with the immune system, and biotherapeutics (e.g., CAR T-cells, Figure 4A) in order to evaluate several key factors (e.g., cancer cell immune evasion) impacting disease prognosis. Their model predicted immunotherapy success probabilities considering immune turnover, antigen escape, and immune adaptive repertoire. A specific prediction was that in acute myeloid leukemia (AML), CAR T-cell therapies have a higher probability of success than cancer vaccines. They also predicted that AML incidence would increase as immunity decreases with age and with use of a chronic immunosuppressor. This simple yet high-quality modeling framework help evaluate the interplay between cancer immunotherapies and cancer progression and immune evasion. The availability of the code and the dataset helps in reproducing the model's predictions. However, assumptions about the model do not include variations of cancer subtype such as antigenicity, T-cell infiltration, and sub-clonal neoantigen landscape. A model including these variables might provide a better prediction of immunotherapy efficacy.

Sahoo et al. (19) developed a simple ODE-based predator-prey mathematical model describing the temporal interplay of glioma and CAR T-cells (Figure 4A) by using *in vitro* assays and *in vivo* patient data, enabling better understanding of the relative impacts of cell proliferation, killing and exhaustion on the outcomes of CAR T-cell therapies. This study is a valuable one, as the model validation and parametrization were performed using both *in vitro* cell-killing assays and *in vivo* clinical trial data from a glioblastoma multiforme (GBM) patient. Use of a simple mathematical model built using knowledge of CAR T was helpful in this application to describe and understand CAR T-cell therapy dynamics. The author's mathematical analysis showed that the model can simulate three scenarios: (1) successful CAR T-cell treatment, (2) CAR T-cell treatment failure, and (3) pseudo-failure or pseudo-response (i.e., model-predicted coexistence of CAR T and target cells). The pseudo-failure or pseudo-response prediction is consistent with clinical observations in which the cancer initially progressed during therapy before finally responding. The model results suggest that the balance between proliferation and exhaustion of CAR T-cells may contribute more to the treatment success or failure than the rate of CAR T-cell-mediated cancer cell killing. Nevertheless, their model is a variation of the classic Lotka-Volterra predator-prey model. It only captured the simple dynamics observed from the data in their experiment, and parameter estimates are uniquely determined by their experiment. This model may not be applicable in other *in vitro* assay conditions. Moreover, the change in the electrical impedance to non-invasively quantify the adherent cell densities, measured as cell index, cannot differentiate cell detachment from cell killing, and this system is not a direct measurement of CAR T-cell dynamics. The intracavitary and intraventricular injections potentially result in spatially heterogeneous densities of CAR T-cells. Therefore, using the modeling assumption of well-mixed cell populations and tumor spheroids might not be easily justified. Application of spatial methods can be a more suitable modeling direction. Last, this model does not include cytokines, stromal cells, or additional immune cells (e.g., myeloid cells which contribute to CAR T-cell activity *in vivo*).

Through the lens of mouse models, de Jesus Rodrigues and colleagues (13) developed a three-compartment ODE-based mathematical model describing the temporal interplay of mouse cancer cells, CAR T-cells and memory T-cells (Figure 4C) by using *in vivo* data in order to better understand the relative impacts of cell proliferation, killing, and tumor immunosuppressive environment on the outcomes of CAR T-cell therapies. Their proposed model can represent the three phases of the therapy: tumor elimination, equilibrium, and escape. Using their model, they have identified that the success of immunotherapy is closely associated with the tumor growth rate, CAR T-cell inhibition and mostly CAR T-cell proliferation, which is in agreement with predictions made by Sahoo et al. (19). It also provides an *in-silico* tool for assessing tumor burden-dependent CAR T-cell dosing, CAR T-cell infusion protocols, and immunosuppressive tumor mechanism. Their model is built not only on known biological mechanisms (dynamics of the tumor, CAR T effector, and memory T-cells), but also on mouse data (39, 40) with tumor targets as HDLM-2 and RAJI cell lines. In general, their model can be adapted to different treatment and tumor scenarios and can help predict therapy efficacy. This model is one of the few models which have included the long-term immunological memory T-cells. On the other hand, this model does not include cytokines and concurrent therapeutics such as tocilizumab, corticosteroids, or checkpoint inhibitors and parametrization was performed on mouse model data.

Kimmel et al. 2019 (15) developed an ODE-based mathematical model that describes the interplay of normal memory T-cells, memory CAR T-cells, effector CAR T-cells and malignant $CD19^+$ B cells (Figure 4C), in order to better understand the relative impacts of cell proliferation and killing of malignant cells on overall CAR T therapy efficacy. To model and predict a patient's "cured state" with tumor cells dwindling in number, authors have developed a hybrid model where the tumor cell population is stochastically modelled if there are less than 100 malignant cells. On the other hand, if the malignant cell count is above that threshold, a deterministic model was used as is the case for all other cell populations in the model. With their model, they predict that clinical interventions (e.g. lymphodepleting chemotherapy) prior to CAR T therapy could improve the duration and likelihood of durable immune responses against a tumor mass. The model also predicts that an increase in memory CAR T-cell fractions (i.e. fraction of $CCR7^+$ cells) beyond 80%, leads to lower progression-free survival (PFS) rates. This mathematical model is well described and easy to understand. One limitation, however, is the lack of clinical trial validation, even though the parametrization was performed on clinical trial data. The focus is on CAR T-cell therapy efficacy and cytokines are not included in this model. Code for the simulations is not available making replication of the results challenging. Finally, the phenomenon of immune evasion of cancer cells was not considered.

A mechanism-based translational Pharmacokinetics-Pharmacodynamics (PK-PD) model developed by Singh et al. (41) integrated key drug-specific and system-specific parameters into a quantitative framework in order to understand the PK-PD determinants of CAR T-cells. Authors have employed a stepwise, bottom-up approach. In the first step, a cell-level PD model was developed integrating the effects of CAR affinity, CAR-densities, antigen densities, and Effector cells : Target cells (E:T) ratios (to compute the rate and determine the extent of saturable tumor cell killing), as well as CAR T expansion, cytokine release, and data from a comprehensive set of *in vitro* experiments were used to account for the dynamic E:T ratio due to T-cell expansion, key drug-specific determinants like CAR-affinity, CAR-density and system-specific determinants like antigen density. This cell-level model captured the quantitative impact of the drug-specific parameters on the overall CAR T-cell activities *in vitro*. In the second step, the authors developed a Physiologically-based Pharmacokinetic Model (PBPK) model to characterize the biodistribution of untransduced T-cells, anti-EGFR CAR T-cells, and anti-CD19 CAR T-cells in all major tissues, including spleen, liver, and lungs. In the third and final step, an integrated PBPK-PD model was developed to characterize the rapid expansion of CAR T-cells in blood and the observed inhibition of tumor growth. Using this translational modeling framework, based on overall comparison of potency parameters across different CAR constructs, authors established an *in vitro* to *in vivo* correlation, and that the *in vitro* potency values were consistently 10–20-fold higher than *in vivo* potency values. These model simulations suggested that CAR T-cells may have a steep dose-exposure relationship, and the apparent $C_{max}$ upon CAR T-cell expansion in blood may be more sensitive to patient tumor burden than to CAR T dose levels. Simulation results also suggested that upon formation of threshold CAR-target complexes per tumor cell, there is increased expansion of CAR T-cells, leading to increased tumor cell eradication. The model simulations were performed using Stochastic Approximation Expectation Maximization (SAEM) algorithm of Monolix version 8 (Lixoft) and the authors have made the model code available in their paper. This comprehensive PBPK-PD model developed using *in vitro* and *in vivo* data can be used to evaluate efficacy/safety in anti-EGFR, anti-HER2, anti-BCMA, and anti-CD19 CAR T-cells. This model can help establish the unmet PK-PD relationships for CAR T-cells. Also having the cell-level information incorporated into the model can help identify optimal CAR T characteristics like CAR-affinity and CAR-density, which will facilitate the selection of lead CAR T candidate at the discovery stage. Moreover, the PBPK model developed will be very useful to study the disposition of CAR T-cells targeted against the solid tumors and also hematological malignancies, where majority of the anti-CD19 CAR T-cells are found in the tissues than in peripheral blood (42, 43), thus facilitating better understanding of CAR T kinetics. However, the model was not validated on clinical data, and by itself is limited by the value of its training *in vitro* and *in vivo* data. With the availability of more relevant clinical data, this model would enable optimal CAR design and development.

The most frequently-seen adverse effect of CAR T-cell therapy is CRS; however, little is known about the underlying biology of this syndrome. An increased understanding could help to better predict risk and improve mitigation strategies so the associated serious complications can be prevented. Below, we have reviewed studies that included cytokines in their model.

Two studies conducted by Teachey et al. (21) and Hopkins et al. (14) have identified cytokines associated with CRS and the cytokine inhibitor targets to alleviate the problem. Teachey et al. measured cytokines and clinical biomarkers in 51 patients, including 39 children (5–23 years) and 12 adults (25–72 years) with r/r ALL treated with tisa-cel, and identified that the peak levels of 24 cytokines in the first month after CAR T-cell infusion were associated with sCRS (21). Although these cytokines were associated with CRS, their levels could not predict the patients who would develop CRS. The authors developed and analyzed 16 predictive models using either regression or decision tree modeling and were able to predict which patients developed grade 4–5 CRS. Using their top regression model, they accurately predicted which patients would develop sCRS using a signature composed of three cytokines, including IFNγ, sgp130 and sIL1RA. These predictions could help in early clinical interventions, thus decreasing morbidity and mortality in CAR T-treated patients. Teachey et al. tested the predictive accuracy of their models both with cytokines alone and cytokines along with initial tumor burden measured from bone marrow aspirates of pediatric cohorts immediately prior to CAR T-cell infusion. They found that incorporating the initial tumor burden into their regression model did not increase predictive accuracy; however, the tumor burden was identified as an important predictive variance in their decision tree model. Model accuracy was validated using 12 additional pediatric patients. The authors also characterized the effects of tocilizumab on CRS, showing that CRS is mediated by trans-IL-6 signaling, which is abrogated after tocilizumab treatment. The findings in this study provide a step forward in predicting severity of CRS and the general cytokine dynamics, even though patient-level cytokine kinetic data was not made available to the readers.

Hopkins et al. (14), on the other hand, used a set of ODEs to better understand the inhibitor/inductor mechanisms of nine important cytokines in CRS and to propose optimized CRS treatment strategies. However, this model cannot be readily applicable to a cell therapy such as CAR T where the cells are alive and proliferating, unlike the cytokine network studied in this article.

Although the CAR T-cell dose, tumor burden, and T-cell expansion kinetics were thought to be associated with the severity of CRS, the quantitative relationship between these factors and occurrence of CRS had not been studied until recently. Mostolizadeh et al. (16) developed an ODE-based mathematical model including CAR T-cells, healthy and cancerous B cells, other circulating lymphocytes and the cytokines in the blood, for modeling CAR T therapy efficacy as well as safety (Figure 4D). They analyzed the model to identify dosage and patient conditions that would result in the optimal therapeutic outcome. The model was analyzed with respect to its equilibrium points, which gives various outcomes of the CAR T-cell therapy, depending on the initial conditions. Authors applied optimal control theory to their model, with one controller for the dose of CAR T-cells injected, and another for tocilizumab. They have determined an optimal dosage of anti-cytokine drug and CAR T-cells with respect to safety and efficacy. However, the model was limited by a lack of clinical and experimental validation and was not able to determine the relationship between tumor burden and CAR T-cell proliferation. Healthy and malignant B cells were killed at different rates by CAR T-cells, and no explanation was provided to justify these rates.

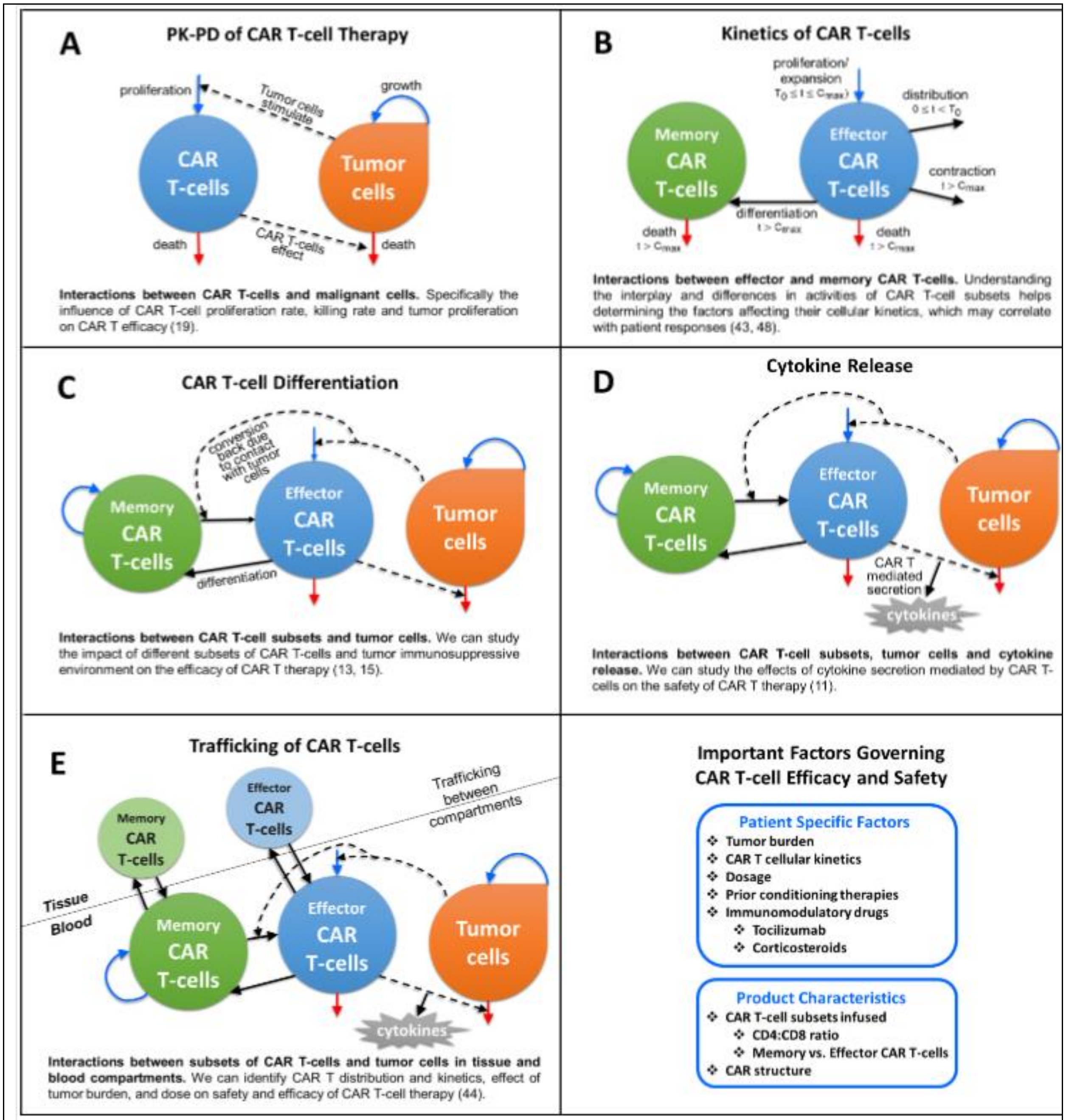

**Figure 4. Schematic representation of key CAR T-cell interactions and potential factors affecting product safety and efficacy.**

In 2019, Hardiansyah and Ng (44) established the first Quantitative Systems Pharmacology (QSP) model that used clinical data to quantify the complex relationships among CAR T doses, disease burden, and cytokines in human subjects and to gain relevant insights into the determinants of clinical toxicity/efficacy of CAR T-cell therapy. This model included CAR T-cell subsets (i.e., effector and memory in peripheral blood and tissues), B cells, and inflammatory cytokine (IL-6, IL-10 and IFNγ) components (Figure 4E). It predicted that the expansion of CAR T-cells and the elimination of B cells are more correlated with disease burden than the administered CAR T dose. The model also predicts that the infused CAR T-cell dose is not corelated with the severity of CRS. This is the first mechanistic model of its kind to study the relationship between CAR T dose, tumor burden, and CRS using clinical data. It can be a good prototype on which to build a more comprehensive CAR T model and can be reproduced using the ODE published in the paper; even though, data from only two patients was used to develop the model.

Another ODE-based mathematical model was developed by Hanson et al. (11, 18) that describes the interplay between malignant CD19$^+$ B cells and CAR T-cells in order to better understand the relative impacts of cell proliferation, killing of malignant cells, and cytokine secretion rates on overall CAR T therapy outcomes (Figure 4D). Using this model, an *in silico* randomized controlled trial (sRCT) was conducted to help understand how inter-patient variation can give rise to a range of clinical responses observed in CAR T-cell therapy and to develop treatment protocols possibly having a higher likelihood of success in different patient populations. These *in silico* trials identified a patient subgroup which might not have been ideal candidates for CAR T-cell therapy. They also demonstrated mechanistically that small differences in initial tumor burden and other patient-specific parameters can result in large differences in therapeutic outcomes and showed correlation between toxicity and initial tumor burden but not with CAR T-cell dose which is in agreement with the clinical trial data (45). Also, when no memory cells were in infused CAR T-cells, toxicity is lower on average, but could result in earlier malignancy relapse. However, the predictions of the model were not validated using clinical trial data, and only a single cytokine was modeled.

In summary, the models reviewed in this section (9-11, 13-16, 18, 19, 21, 41, 44) strive to analyze the complex interactions between the tumor and immune cells including CAR T-cells and/or the cytokines (Figure 4) using mathematical modeling. These models can be essentially applied to study the therapeutic outcomes at different patient conditions and doses. However, the models reviewed in this section for the most part have limited input of clinical data. Although having adequate clinical data is a challenge for modeling in general, it is particularly so for CAR T-cell modeling, as the CAR T-cell therapy field is still in its infancy and there is a lack of consensus on key factors driving therapeutic efficacy and safety.

## Pharmacokinetic models studying CAR T kinetics in patient cohorts

Studies included in this section used population data instrumental in understanding the disparities in CAR T-cell therapy outcomes, in identifying the covariates that influence variability, and in quantitatively estimating the variability commonly seen in CAR T-cell therapy. The typical pharmacokinetic components, including distribution, metabolism and excretion, are not directly applicable to CAR T-cells, as CAR T-cells proliferate *in vivo* after they are infused. Hence, studying the cellular kinetics of CAR T-cells refers to their *in vivo* characterization (46). Several studies have attempted to identify the relationships between the cellular kinetics of CAR T-cells and their efficacy and safety and the factors affecting CAR T kinetics.

Unlike any other small drug molecule, CAR T-cell is a living drug, which undergoes rapid expansion after infusion, several folds beyond its dose and demonstrate long-term persistence based on the presence of its target antigen and other factors. CAR T-cells exhibit multiphasic kinetics with initial exponential expansion up to $T_{max}$ followed by biexponential contraction and a gradual decrease of CAR T-cells that can persist in the body for months or years. Stein and colleagues (43) have developed the first nonlinear mixed-effect semi-mechanistic model which has captured the three distinct cellular kinetic phases (Figure 4B), to investigate the different peak tisa-cel levels and the impact of tocilizumab and corticosteroids on the rate of CAR T-cell expansion. The study used data from two phase-2 clinical studies, ELIANA and ENSIGN, including pediatric and young adult r/r B cell ALL patients. This population PK model is the first of its kind to study the random effects, and the patient-specific covariates. Six random effects were included on the estimated model parameters corresponding to the rate constants of different phases of CAR T kinetics. Several categorical and continuous covariates were explored, and their boot-strapping results showed there was no statistically significant impact of any of these covariates on peak plasma concentration ($C_{max}$). Doubling time, initial decline half-life, and the terminal half-life of tisa-cel were reported to be 0.78, 4.3, and 220 days, respectively. The authors concluded that there is no impact of tocilizumab or corticosteroids on CAR T-cell expansion rate, as observed by Mueller et al. (46, 47). The model estimated $C_{max}$ was two-fold higher in patients who received tocilizumab. Those patients might have had higher CAR T-cell expansion which would lead to CRS since CRS is an on-target toxicity. Although some patients received multiple doses of tocilizumab and corticosteroids, only impact of the first dose was modeled and an interaction term between tocilizumab and corticosteroids was not included in the model. Furthermore, the effect of corticosteroids at higher doses given alone or before tocilizumab, were not modeled. Yet, this model can be adapted to characterize the expansion and persistence of different types of CAR T-cells manufactured for different indications, as in the case of Liu et al. (48) where authors studied CAR T-cell kinetics in patients with different tumor types using Stein's model with few modifications.

Recently, Liu et al. (48), using Stein's model (43) and De Boer's ODE-based simple longitudinal mechanistic model of $CD4^+$ and $CD8^+$ cell subset counts (49) with few modifications, developed a cellular kinetic model to characterize multiphasic kinetics of CAR T-cells (Figure 4B), including the initial distribution phase seen in lymphoma patients. The study included 217 patients from seven 4-1BB CAR T-cell therapy clinical trials, targeting different tumor antigens in hematological malignancies (acute lymphocytic leukemia (ALL), chronic lymphocytic leukemia (CLL), multiple myeloma (MM), lymphomas (diffuse large B-cell lymphoma (DLBCL)) and solid tumors (non-small-cell lung cancer (NSCLC), glioblastoma (GBM)). This is the first study where interindividual variability (IIV) is identified within multiple clinical trials and compared across these clinical trials.

Quantification of IIV aids in development of safe and efficacious dosing strategies for CAR T-cell therapy. Liu and colleague's study is valuable as it can be used to identify the patient specific, product specific and tumor type dependent factors that correlate with CAR T-cell kinetics and patient responses. Authors have compared CAR T kinetics across response groups, tumor types and analyzed the correlation of CAR T-cell kinetics with dose, tumor burden, CAR T-cell composition, patient age, and lymphodepletion. The model has well captured different phases of CAR T cellular kinetics and the model parameters were estimated with acceptable precision and variability. Their analysis shows that, the proliferation capacity ($C_{max}$) is higher in responders compared to non-responders with ALL (adult), CLL, MM and NSCLC. This response was partly due to the higher proliferation rate of CAR T-cells, and relatively slower contraction rate, however no statistical significance was seen. In hematological malignancies (ALL/CLL/MM) the proliferation capacity ($C_{max}$) and proliferation duration were significantly higher than in lymphoma and solid tumors (NSCLC and GBM). The reason for this outcome could be the poor antigen accessibility in those compared to hematological cancers. Also, the variability seen in the proliferation rate amongst the two solid tumor trials (NSCLC and GBM) may be attributed to the difference in tissue accessibility and the delivery rate of CAR T-cells.

The comparisons between the dose (ranging $10^7$- $10^9$ cells/patient) and patient responses was not shown to be statistically significant in CLL, MM, GBM, and NSCLC patients in Liu et al.'s work. The pretreatment tumor burden in MM and CLL patients was not statistically correlated with patient responses; however, in these studies tumor antigen biomarker was quantified only from the blood. The CD4:CD8 ratios in MM and NSCLC showed negative correlation with proliferation capacity ($C_{max}$) and proliferation rate, though it is not statistically significant. In this study, patient age was found to be associated with immune response, pediatric ALL patients showed lower proliferation rate but a lower contraction rate, higher memory differentiation and lower memory death rate. Patients in the complete response/partial response group appeared to be younger than patients in progressive disease/no response group, although there is no statistical significance. Using MM trial, analysis of two cohorts of patients given same CAR T-cell dose with or without lymphodepletion, showed that patients who underwent lymphodepletion have higher CAR T proliferation capacity ($C_{max}$). In NSCLC trial, this correlation was not found, however lymphodepletion was only applied to patients with larger tumor size.

Additionally, in Liu et al. correlation analysis among the model parameters showed that the higher CAR T proliferation capacity ($C_{max}$) was associated with faster proliferation rate and longer proliferation duration. The CAR T contraction rate is slightly negatively correlated with proliferation rate and proliferation capacity. To explore the IIV and interstudy variability, the team has simulated virtual population of 1000, built using trial specific model parameters along with IIVs for both effector CAR T-cells and memory CAR T-cells. Responders in ALL and MM trails showed higher peak of effector CAR T-cells, higher formation of memory CAR T-cells and slower contraction rate of effector CAR T-cells than non-responders. There is considerable heterogeneity in effector and memory CAR T-cells across the trials as well as among the population in each trial, shown by their simulations. However, these comparisons originated from clinical trials studying CAR T-cell therapies targeting different antigens, have different product manufacturing strategy and different patient characteristics. As a result, generalization of these findings to CAR T-cell therapies as a whole may not be possible and more clinical studies need to be performed and long-term longitudinal data of CAR T-cells need to be made available to make better comparisons and find statistically significant correlations between CAR T kinetics and various potential factors affecting it.

In summary, the studies reviewed in this section are empirical PK models (43, 48) and these models quantified the interindividual and/or intertrial variabilities in CAR T-cell kinetics and this gives an idea of the biological determinants of the IIV in CAR T-cell kinetics. These models explored correlations between therapeutic prognostic factors such as, lymphodepletion, infused CAR T-cell composition, initial tumor burden and the therapeutic efficacy. However, these correlations may not be extrapolated to newer generation of CAR T-cell therapies because of the underlying product specific factors.

## Statistical survival models for patients undergoing CAR T-cell therapy

To estimate the long-term overall survival of patients with DLBCL treated with axi-cel, Bansal et al. (8) compared the standard parametric models to mixture cure models. This study demonstrated that the traditional parametric survival models can underestimate the overall survival with CAR T-cell therapy because there is a substantial variability in the extent and timing of patient responses to CAR T. Although both the standard and mixture cure models showed successful fitting to the currently available data (up to about two years' follow-up), the two models differed substantially in their extrapolated survival outcomes beyond two years. For example, the 10-year survival probability predicted by the mixture cure models was 0.5, while standard models predicted probability of less than 0.2. As this is a statistical model, the model parameters were estimated using clinical trial data. However, the model forecasts currently cannot be validated due to small sample size and short follow-up periods in the clinical trials of CAR T-cell therapies, and long-term survival/efficacy/safety of the patients are unknown. Also, they focused only on one study, which raised concerns that the predictions may not be easily reproduced. The code for their model was not made available, but implementation of the model using the equations in the article is straightforward.

Grant et al. (50) provided a detailed analysis on the fitness of cure models when predicting patient survival data. The authors simulated a case study of virtual patients based on actual disease progression rates reported in the literature. They simulated CAR T-cell therapy survival data and analyzed the goodness of fit of non-mixture and mixture cure models to evaluate the usefulness of cure modeling in cancer survival prediction. The authors showed how real-life data could affect survival predictions made by cure modeling. Particularly, their results suggest that cure modeling techniques should not be used if the real-life survival data is immature (i.e., early phase in a clinical trial with insufficient follow-up). On the other hand, unnecessarily long follow-up periods in a dataset may involve age-related mortality as a potential confounder for the model. The optimal time point to use cure modeling is disease-specific, and clinician input is necessary in finding that time point. However, the conclusions of this study are specific to the data generated in it, and applying the techniques used in this paper to immature data could result in inaccuracy of the cure fraction.

In summary, articles reviewed in this section (8, 50) analyzed the suitability of cure models in predicting the long-term survival of CAR T-cell therapy treated patients. These studies would have made better long-term predictions of overall survival, if Long term follow-up data was available.

## Pharmacoeconomic models

Often, the information pertaining to long-term benefits and risks of CAR T-cell therapies is limited, which makes determining cost-effectiveness of CAR T-cell therapy difficult. More importantly, the clinical trials that led to CAR T-cell therapy approvals, and many ongoing trials, are single-arm trials. Pharmacoeconomic models are important evidence-based tools in forecasting and evaluation of CAR T-cell therapy outcomes and are a means of justifying therapeutic cost-effectiveness. Here, we summarize the pharmacoeconomic studies developed to determine the cost-effectiveness of CAR T-cell therapy and their advantages and disadvantages.

Using Markov cure/survival modeling, Lin et al. (51) determined the cost-effectiveness of tisa-cel for pediatric B cell ALL patients versus the standard line of therapy. This model provided a sequence of important milestones in the outcomes of pediatric r/r ALL patient populations that receive CAR T-cell therapy or chemotherapies using detailed patient and product attributes informed by clinical trials. The cost-effectiveness analysis of tisa-cel was conducted for multiple scenarios using a well-informed model. Based on their analysis, the authors recommended a reduction in the price of tisa-cel (from $475,000) to $200,000 or $350,000, allowing the therapy to meet a $100,000/quality-adjusted life year (QALY) or $150,000/QALY willingness-to-pay threshold in all outcome scenarios. One limitation similar to other models is lack of high-quality long-term clinical outcomes data. Another issue was that all trials for relapsed or refractory pediatric ALL were single-arm studies, which limit a direct comparison between tisa-cel and standard-line therapy. Additionally, the authors did not account for tisa-cel's non-health care benefits, such as future productivity, which may be substantial given the young age at which patients are treated.

Using a multi-state survival model, Furzer et al. (52), using data of 192 patients from three pooled clinical trials and 118 patients from the cancer registry of Canada, quantified the value of tisa-cel compared with current standard care for eligible pediatric patients with relapsed ALL. Even though the authors were unable to use a randomized clinical trial dataset due to unavailability, they incorporated sensitivity analyses to demonstrate the robustness of their cost utility analytic results. They found an incremental cost per QALY gain of tisa-cel over standard care (US $53,933–$213,453) based on assumed cure rates of 40%–10%. Like similar models, this type of modeling would provide higher estimating precision as longer-term outcomes data becomes available.

Using a very detailed microsimulation cure/survival modeling framework, Sarkar et al. (53) determined the cost effectiveness of tisa-cel for pediatric B cell ALL patients with regard to the standard line of therapy. The model produced outcomes of progression, survival, and toxicity similar to estimates in the literature. Cost-effect analysis was most sensitive to assumptions of long-term CAR T survival, proportion of CAR T patients achieving complete remission, and health utility of post-treatment patients. Their probabilistic sensitivity analysis found that CAR T was cost-effective in 94.8% of iterations at a willingness to pay $100,000/QALY. Also, if one-year survival was decreased to 57.8%, then CAR T was predicted to no longer be cost-effective. This type of modeling has limitations. First, due to lack of long-term data on survival, cost, role of hematopoietic stem cell transplantation (HSCT) after CAR T, and the complications involved, the findings in this article could vary as more data becomes available. Second, the quality of data used to inform this model of cost-effectiveness should be from a randomized phase 3 trial comparing CAR T therapy to the standard of care. This research was based on the available data without information about how patients not responding to CAR T would respond to chemotherapy afterward, as the authors assumed that patients had similar responses to chemotherapy as those who initially did not receive CAR T therapy.

Whittington and colleagues (12) estimated the long-term survival of r/r leukemia patients younger than 25 who used tisa-cel, and then estimated the actual value of tisa-cel based on long-term survival rates using a mixture cure model and a decision analytic model including a short-term decision tree and a long-term semi-Markov partitioned survival model. They parametrized the model using published literature on therapeutic (CAR T and clofarabine) safety and efficacy outcomes and the pertinent economic parameters. The study suggests that tisa-cel in pediatric patients with B-ALL provides clinical benefits in quality-adjusted and overall survival compared with clofarabine, and that tisa-cel seems to be priced in alignment with its clinical benefits.

In their recent study, Whittington and colleagues (54), estimated the long-term survival of adult r/r non-Hodgkin lymphoma patients who used axi-cel, and then estimated the actual value of axi-cel based on long-term survival rates with respect to standard chemotherapy. Treatment with axi-cel appeared to be associated with positive, yet uncertain, gains in survival compared with chemotherapy, and more cost-effectiveness because of better long-term survival. However, the predictions from this model cannot currently be validated, as not enough patients with sufficiently long follow-up periods are available. Also, there is no data from randomized controlled trials directly comparing CAR T-cell therapies against chemotherapies. However, when building and parametrizing their model, the authors assumed that their data came from a single RCT. Plus, the authors compared CAR against a single alternative treatment only and no other context-relevant therapy (e.g., blinatumomab in leukemia). Because of the uncertainty in long-term survival extrapolations and corresponding assumptions, the authors claimed that it is important to generate and present the results from multiple potential survival models that have differing but plausible assumptions. Therefore, they fitted five different survival models to the published survival curves, by which the variation in long-term survival assumptions were captured and a range of long-term survival estimates were generated.

Roth et al. (55) using cure/survival modeling, determined the cost-effectiveness of axi-cel for U.S. adult r/r B cell lymphoma patients compared to the standard line of therapy. The authors determined that the cost of axi-cel was $58,146 per QALY gained. This prediction was most sensitive to the long-term remission rate, drug price discounts, and axi-cel price. The current data of ZUMA-1 is limited at a median patient follow-up of 15.4 months, and longer-term outcomes should be considered for model validity. They modeled the R-DHAP regimen as a comparator treatment strategy, a common guideline-recommended treatment, but alternative regimens could be considered.

In summary, pharmacoeconomic models reviewed in this section (12, 51-55), evaluated the efficacy and safety of CAR T-cell therapy and its therapeutic cost-effectiveness with respect to standard line of therapy. Since, long-term clinical outcomes are not readily available from clinical trials with limited follow up periods, results of these model simulations provide long-term projections on product benefit and risk, along with treatment costs.

**Table 1.** A Summary of the CAR T-cell Therapy modeling and simulation articles. MLE: Maximum Likelihood Estimation, ODE: Ordinary Differential Equations, CAR: Chimeric Antigen Receptor, PLS: Partial Least Squares, MC: Monte Carlo, MCMC: Markov Chain Monte Carlo, QSP: Quantitative Systems Pharmacology, PK: Pharmacokinetics, PBPK: Physiologically-based Pharmacokinetic Model, PD: Pharmacodynamic, PSO: Particle Swarm Optimization, eFAST: Extended Fourier Amplitude Sensitivity Test, SRCT: Simulated Randomized Control Trial, PIA: Parameter Identifiability Analysis.

| Article | Model Scope | CAR Construct Focus | Model Type and Algorithms Used | Simulation and Analysis Tools | Data Provided | Code Provided | Model Fitting and Validation |
|---|---|---|---|---|---|---|---|
| **Models of CAR T-Cell Activation Signaling Cascade** | | | | | | | |
| Cess & Finley 2020 and Rohrs et al. 2018/2019 (34-36) | Single cell level with intracellular signaling detail | CD28-CD3ζ CAR | ODE-based Mechanistic, MC Sampling, PSO, eFAST, PLS | MATLAB, Prism, BioNetGen | No | Yes | Fitted/validated with *in vitro* data |
| Harris et al. 2018 (29) | Single cell level with intracellular signaling detail | CD3ζ and CD28-CD3ζ CAR | ODE-based Mechanistic | Not reported | No | No | No |
| **Models used to evaluate and predict CAR T-cell therapy safety and/or efficacy** | | | | | | | |
| Finney et al. 2019 (9) | Population of cells | CD19CAR-T2A-EGFRt | Statistical, Classification, and Regression Tree analysis | R, SAS, Prism | No | No | Fitted with clinical trial data |
| de Jesus et al. 2019 (13) | Population of cells | Anti-CD19 and anti-CD123 CAR | ODE-based Mechanistic, MCMC | QUESO Library | No | No | Fitted with *in vivo* mouse data |
| George and Levine 2018 (10) | Population of cells and patients | No | Therapeutic outcome modeling, Stochastic and deterministic variants | MATLAB | Yes | Yes | No |
| Hanson et al. 2016-2019 (11, 18) | Population of cytokines, cells and cohort of patients | No | QSP, ODE-based Mechanistic | MATLAB | No | No | No |
| Hardiansyah and Ng 2019 (44) | Population of cytokines, cells and cohort of patients | CD3ζ-4-1BB | QSP, ODE-based Mechanistic | MATLAB Simbiology and SAAMII | No | No | Fitted with patient-level clinical trial data and validated on patient data |
| Hopkins et al. 2019 (14) | Cellular Cells, cytokines | No | ODE-based Mechanistic | Not reported | No | No | No |
| Kimmel et al. 2019 (15) | Population of cells and cohort of patients | CD28-CD3ζ CAR | Hybrid (stochastic and deterministic), Gillespie Algorithm | Julia | No | No | Fitted with ZUMA-1 trial data |
| Mostolizadeh et al. 2018 (16) | Population of cytokines and cells | No | ODE-based Mechanistic | MATLAB | No | No | No |
| Sahoo et al. 2020 (19) | Population of cells | IL13BBζ | PDE/ODE-based Mechanistic | MATLAB, PIA | Yes | No | Fitted with *in vitro* data, validated on *in vivo* MRI data |
| Teachey et al. 2016 (21) | Cytokine levels in cohorts of patients | CD3ζ-4-1BB | Statistical, Regression and Decision Tree analysis | R and SAS | Yes | No | Fitted with clinical trial data |
| Singh et al. 2020 (41) | Population of cells and cytokines from patient cohorts | Anti-EGFR, Anti-HER2, Anti-BCMA, Anti-CD19 | ODE-based Mechanistic, PBPK/PD | Monolix SAEM | No | Yes | Fitted with in vitro and in vivo data. Not validated |
| **Pharmacokinetic models studying CAR T-cell kinetics in patient cohorts** | | | | | | | |
| Stein et al. 2019 (43) | Population of cells | CD3ζ-4-1BB | Statistical/empirical PK | Monolix, R, and MATLAB | Yes | Yes | Fitted with clinical trial data |
| Liu et al. 2020 (48) | Population of patients | CD3ζ-4-1BB | Empirical PK | Monolix and R | Yes | Yes | Fitted with clinical trial data |
| **Statistical survival models for patients undergoing CAR T-cell therapy** | | | | | | | |
| Bansal et al. 2019 (8) | Population of patients | CD3ζ-CD28 | Statistical cure modeling, MLE | Strata | No | No | Fitted with clinical trial data |
| Grant et al. 2019 (50) | Population of patients | CD3ζ-CD28 and CD3ζ-4-1BB | Statistical cure modeling | R flexsurvcure | Yes | No | Fitted with clinical trial data |
| **Pharmacoeconomic models** | | | | | | | |
| Whittington et al. 2018-2019 (12, 54) | Population of patients | CD3ζ-CD28 and CD3ζ-4-1BB | Pharmacoeconomics, Statistical cure modeling, Decision Trees | MS Excel, R | No | No | Fitted with clinical trial data |
| Furzer et al. 2020 (52) | Population of patients | CD3ζ-4-1BB | Pharmacoeconomics, Cost-utility | R | No | No | Fitted with clinical trial data |
| Lin et al. 2018 (51) | Population of patients | CD3ζ-4-1BB | Pharmacoeconomics, Cost-utility, Markov cure/survival | TreeAge Pro, R | No | No | Fitted with clinical trial data |
| Roth et al. 2018 (55) | Population of patients | CD3ζ-CD28 | Pharmacoeconomics, Cost-utility, Multi-state survival | MS Excel, SAS | No | No | Fitted with clinical trial data |
| Sarkar et al. 2019 (53) | Population of patients | CD3ζ-4-1BB | Pharmacoeconomics, Cost-utility, Multi-state survival | TreeAge Pro | No | No | Fitted with clinical trial data |

# DISCUSSION

Mathematical and computational models facilitate the ability to quantitatively bridge the gap between data gathering and mechanism testing (56). Computational models can provide a set of analytical and numerical tools, complementary to laboratory experimentation, for understanding the underlying mechanisms that drive the observed biological phenomena across multiple scales, from the cellular-molecular level up to human population groups. They, therefore, provide new insights on key interactions within a system and demonstrate whether a hypothesized mechanism explains observed phenomenon (56). In addition, these models have the advantage of avoiding the difficult challenge of performing laboratory experiments that are currently not viable or ethical in a physical system (56). As an example, computational modeling of immune system dynamics in response to CAR T-cell therapy provides novel insights into the complex interactions between the human immune components and CAR T-cells; helping to explain the existing observations, predict potential outcomes, and generate hypotheses that can be tested *in vitro*, or *in vivo* (57). It is well understood that all models are inherently limited, but they are useful tools if appropriately constructed and validated and satisfy the good modeling practices discussed extensively elsewhere (58-60).

## What are the added values of CAR T modeling and simulation efforts?

Regulatory approval of CAR T-cell therapies and demonstrated clinical effectiveness open new avenues for novel cellular immunotherapies. However, current CAR T-cell therapies face substantial challenges, including product manufacturing issues, life-threatening adverse events, and high cost. As we continue to better understand the factors triggering therapeutic safety and efficacy issues, scientists have applied computational modeling and simulation to help link these factors to clinical outcomes.

Several computational models have been developed to investigate the aspects of CAR T-cell therapy, including mechanistic ODE-based mathematical models, as simple as the well-known Lotka-Volterra (predator-prey) model, to study the interplay between the CAR T-cells, tumor cells, and cytokines (17, 44), and to understand the mechanisms of CAR activation with models at the cellular level (34, 35). Also, PK models have been used to study CAR T-cell proliferation (48) and the effect of tocilizumab (43). Moreover, machine learning approaches have been implemented to identify biomarkers of CRS (21); pharmacoeconomic models have been proposed for evaluating cost-effectiveness of CAR T-cell therapy (51-53, 55), and multiscale PK-PD models (41) have been developed to understand and determine the factors responsible for therapeutic outcome.

Cellular level models have been used to help evaluate and identify the optimal CAR T characteristics, including signaling domains, CAR affinity, CAR density, and antigen density, to select the lead CAR T candidate at the discovery stage, or to understand the intracellular signaling mechanisms responsible for the activation and expansion of CAR T-cells (29, 35), such studies can help identify the underlying factors for therapy failure, can save time and resources, and may help come up with better clinical protocols.

Although CAR T-cell therapy has shown tremendous efficacy in terms of disease remissions, the variable outcomes in patients must be studied further and should be better understood in order to develop better CARs. Using a simple mathematical model of predator-prey dynamics, Sahoo et al. (19) explored the nonlinear dynamics involved in CAR T-cell therapy and explained the variations in patient-specific responses, even when the same CAR T-cell dose was given, and also proposed the factors important for the success of the therapy. This model also suggested that the CAR T-cell dose can be tailored according to the patient's tumor growth rate and antigen level to maximize therapeutic benefit, and such hypotheses can be tested in *in vitro* or *in vivo* systems to optimize the dose and CAR T-cell treatment regimen in clinical trials for personalized therapies. One of the advantages of modeling is to help reduce the number of clinical trials required to optimize treatment regimen, target patient group, test combination therapy, and to predict the possible adverse effects. Hanson, using an ODE-based mathematical model, generated virtual patients and simulated them to study the quantitative relationship between CAR T-cells, B-cells, and cytokines at different dosage regimens. Such randomized clinical trials may require more time and resources to conduct and may not be practical in a wider population. In such a case, simulated randomized trials of virtual patients would be of immense help in testing the virtual patient cohorts at different doses and different initial tumor burdens to optimize CAR T-cell therapy for positive benefit-risk profiles. Such randomized simulations of virtual patients also help in understanding the interpatient variability and facilitates extrapolation to different populations of interest, including pediatric, gender-based, and disease stage-based (57). PK models (43, 48) have well captured the multiphasic kinetics of CAR T-cells and quantified the inter-individual variability, which can explain the variable therapeutic outcomes in patients to some extent. Impact of other factors such as lymphodepletion, initial tumor burden, and composition of infused CAR T-cells, on the therapeutic efficacy was explored in these studies.

One of the major limitations of CAR T-cell therapy is CRS, and the ability to predict the patients who will likely have severe CRS enables early interventions that would reduce morbidity and mortality. Teachey et al. (21), using a regression modeling approach, predicted which patients would develop severe CRS, which might not otherwise be predicted using standard lab tests, as many of the cytokines peaked after the patients became ill. This study generated a hypothesis that patients who develop CRS develop clinical and biomarker profiles consistent with macrophage activation syndrome. This comparison may help bring new insights into the biology of CRS development and help manage the risk. Another ODE-based model developed by Hopkins et al. (14) helps in understanding inhibitor mechanisms of the cytokines, which in turn helps in optimizing CRS treatment strategies.

The complicated and labor-intensive production of CAR T-cell therapies puts them among the most expensive drugs currently available on the market. Tisa-cel currently is the most expensive cancer therapy, with a single-infusion cost of $475,000 (51) and for axi-cel, a lifetime cost of $552,921 (55). Pharmacoeconomic models are important evidence-based tools in forecasting and evaluation of therapy outcomes and a means of justifying therapeutic cost-effectiveness. Although long-term efficacy data is limited for CAR T-cell therapies, several pharmacoeconomic models were developed to evaluate CAR T-cell cost-effectiveness using Markov cure/survival modeling that have described a positive relationship between cost and quality of life in CAR T-cell-treated patients (12, 51-55).

## Current Gaps in Knowledge and Future of Modeling in CAR T-Cell Therapies

Despite recent advances in CAR T-cell modeling approaches, there are still some gaps on the road to further the development and improve applications of CAR T-cell therapies. For example, there is no clear understanding on the minimum effective dose required, subtypes of infused CAR T-cells, impact of prior therapies, adverse events (AE), impact of co-medications counteracting AE and the source of variability in CAR T-cell therapy outcomes. The CAR T landscape is evolving quickly, with many emerging CARs designed against novel antigen targets, e.g., CD20 or infectious diseases (5). It is not known to what extent current modeling approaches and their conclusions would be extrapolatable to the new generation of CAR T therapies. Also, some of the current model predictions cannot be validated because of small sample sizes, short patient follow-up times and missing randomized clinical trial data, more problematic for statistical cure models and pharmacoeconomic models of CAR T-cell therapies. Regardless of all the progress, there is no clear understanding on what CAR T characteristics and patient characteristics lead to the heterogenous CAR T kinetics and patient responses. In this section we will highlight the major obstacles in the development of CAR T-cell therapies and possible future directions for modeling CAR T-cell therapy.

**Differences in kinetics of small drugs and CAR T-cells:** The physicochemical properties of small drug molecules and the CAR T-cells are different, which not only impacts the pharmacokinetics, but also safety, efficacy, and product manufacturing strategies. Unlike the small drug molecules, infused CAR T-cells can proliferate exponentially to nearly >1000 folds (43) depending on the abundance of the target cells or patient specific (prior therapies) or drug-specific (CAR T-cell composition) factors. The dose-response relationship of CAR T-cell therapies has not been well-characterized, as CAR T-cell expansion, persistence and efficacy are not correlated with the dose administered (47), an essential part of traditional PK-PD models. Moreover, infused T-cells are heterogeneous in phenotype, unlike small drug molecules and the optimally-effective and safe dose might vary depending on the CAR T-cell phenotype composition (61), lymphodepletion chemotherapy (62, 63), baseline tumor burden (46, 64), and disease type (48). Hence, the traditional PK models may not be readily applicable to study the cellular kinetics of CAR T-cells.

The key differences between CAR T-cell kinetics and small drug molecule PK were nicely highlighted by Stein et al. (43). PK models which can capture the multiphasic kinetics of CAR T-cells are being developed (41, 43, 48); however, these models only provide a framework to understand the kinetic behavior of CAR T-cells observed in the patients. Availability of more data, for example better-defined CAR T-cell dose with T-cell subset standardization and longitudinal/cellular kinetic immunophenotyping data sets, will enable further development of these elementary models to account for drug-specific factors impacting CAR T-cell therapy efficacy and safety and will be particularly valuable in model validation. Further, the development of a mechanistic model framing CAR T landscape (i.e. patient specific, and drug-specific factors) is urgently required to aid in the development of CAR T as a personalized therapy.

**CAR T-cell immunogenicity:** Due to their chimeric molecular nature, CAR T-cells have the potential to cause unwanted immunogenicity, which is both a safety and an efficacy concern (65). Most current adoptive immunotherapy clinical trials utilize autologous T-cells, which can be hampered by the poor quality and quantity of T-cells, as well as by the time and expense of manufacturing autologous T-cell products. Thus, using genetically engineered allogeneic "universal" CAR T-cells could circumvent the limitations of using autologous T-cells and could potentially be developed into a next-generation, highly-efficient CAR T-cell therapy. Such off-the-shelf CAR T-cell therapies can be generated from healthy donors to treat multiple patients. The major barrier preventing the successful use of allogeneic T-cells is unwanted immune responses to the allogeneic T-cells by the recipients, as TCRs on allogeneic cells may recognize the alloantigen molecules of the recipient (leading to Graft vs host disease or GVHD) and the expression of HLAs on the surface of allogeneic T-cells may lead to their rapid rejection. Successful prediction of CAR T immunogenicity in individual patients might be an exciting and important research direction for modelers.

**CAR T-cell therapy for solid tumors:** Despite extensive research, CAR T-cell therapy for solid tumors has not been successful yet. The three main barriers to target solid tumors are (1) locating the tumor-associated antigen, (2) actionable antigen binding affinity and (3) surviving the immunosuppressive tumor microenvironment. Moreover, in most of the ongoing clinical trials of CAR T-cell therapies for solid tumors, tumor associated antigens (TAA) are the targets because of the shortage of tumor specific antigens (TSA) and targeting TAA may cause damage to normal cells/tissues when some of the normal cells overexpress the same antigen. In recent years, strategies to overcome these physical barriers, improving CAR T-cell infiltration and antitumor activity are being applied to successfully treat solid tumors with CAR T-cell therapy. Though some of the on-target/off-tumor effects have been taken care of, for example, with the development of EGFRvIII (TSA) targeting CAR T-cells for glioblastoma, there are still a number of barriers to be addressed for CAR T-cell therapy application on solid tumors, which are detailed elsewhere (66-69). Sahoo et al. (19) using a simple ODE-based predator-prey mathematical model showed that the balance between proliferation and exhaustion of CAR T-cells was important for the efficacy. Though this model studied some of the factors impacting the efficacy, models studying the main hurdles (such as tumor recognition, infiltration, antigen binding, and survival in the tumor microenvironment) in applying CAR T-cell therapy to solid tumors are currently not available. Future models studying these factors will be of more importance in overcoming these hurdles for successful application of CAR T-cell therapy for solid tumors.

**Pre-clinical animal models:** Multiple pre-clinical animal models including syngeneic, xenograft, transgenic and humanized mouse models are being used to study CAR T-cell therapy associated adverse events including on-target/off-tumor toxicities, CRS and neurotoxicity. However, because of variability in cross-species reactivity to nonhuman target antigens and/or the lack of the host immune system, most of the animal models often fail to predict potential adverse events in human, leading to false sense of safety (70). Recently, macaques, which have similar immune system as humans are being used to study CAR T-cell therapy associated toxicities; however, because of the ethical and monetary reasons these primate models are used only after extensive validation in mouse models, as a last step before initiating clinical trials. Advancements in breeding transgenic mouse strains, humanized mouse models and using combinations of multiple animal models, would provide more information about CAR T-cell therapy associated safety and efficacy. Although the model developed by de Jesus et al. (13) facilitates identification of key factors in therapeutic efficacy, a more comprehensive mechanistic model with cytokine dynamics would be beneficial in evaluating various intrinsic and extrinsic factors responsible for both safety and efficacy.

**Effect of immunomodulatory drugs:** Tocilizumab and corticosteroids are used to alleviate adverse events attributable to CAR T-cell therapy, and studies have shown that treatment with tocilizumab in patients experiencing sCRS did not impact the CAR T-cell peak levels (43, 47). Hay et al (64) proposed that the early intervention of tocilizumab may help reduce the severity of CRS, since reduction of CAR T-cell dose as a sole strategy to mitigate toxicity may reduce the efficacy (64). Though, recent clinical studies show that the early intervention of tocilizumab does not impact the expansion, persistence and efficacy (71) and preemptive administration of tocilizumab decreases incidence of grade 4 CRS (72), further studies are required to elucidate the optimal timing of tocilizumab administration and its impact on CAR T-cell efficacy (73).

**Patient specific factors/Interindividual variability:** Mueller et al. (46) elucidated the relationships among CRS, tumor burden, cell expansion, and response by characterizing the in vivo cellular kinetics of CAR T-cell therapy across multiple diseases, including pediatric B-ALL, adult ALL, and CLL patients (n=103) that received either a single dose of CTL019 or two to three fractionated doses within the first 28 days, using noncompartmental analysis. Their cellular kinetic analysis showed that the patients with longer CAR T-cell persistence maintained longer event-free survival. Similarly, using data from JULIET phase 2 clinical study (NCT02445248), Awasthi et al. (42) demonstrated that adult patients with r/r DLBCL, with sustained responses showed longer persistence of CAR T-cells compared to non-responders. Mueller et al. (46) reported that the responding patients showed a coefficient of variation (%CV) for $C_{max}$ and AUC from day 0-28 of 167% and 209% respectively, whereas non-responding patients showed even wider variation, with %CV for $C_{max}$ and AUC from day 0-28 of >1000%. These results show that there is a lot of interindividual variability and the varied CAR T-cell expansion and persistence in responders vs non-responders in pediatric B-ALL, CLL and DLBCL patients may be attributed to patient specific factors, such as tumor killing capacity of the T-cells, prior lymphodepletion and infused CAR T-cell composition among others. PK models developed by Stein et al. (43) and Liu et al. (48) well captured the CAR T-cell kinetics and quantified the interindividual variability and intertrial variability, which were also demonstrated by their simulation results. Their findings would help in adjusting the CAR T dose to achieve favorable benefit-risk outcome and in understanding the patient-specific factors driving therapeutic outcome variability. However, such models might have assessed significant covariate effects in more detail, if there were additional clinical data available.

**Discordance in the CAR T-cell literature:** With the increasing number of studies in the field of CAR T-cell therapies, there comes discordance in literature regarding the factors that impact safety and efficacy outcomes of these therapies. Statistical and exploratory analyses are highly important in CAR T-cell therapy model building and simulations, but the discordances found in literature are misleading. We would like to caution the reader that there is a diversity of conclusions from CAR T clinical trial articles for several reasons (74, 75) and this section provides only a small sample of those differences.

Mueller et al. (47) performed the dose-exposure analysis in B-ALL patients treated with tisa-cel and found that dosage is not correlated with peak plasma concentration ($C_{max}$) or the 28-day area under the concentration curve (AUC0-28D). Similarly, correlation analysis of tisa-cel dose and exposure (i.e., AUC) in DLBCL patients performed by Awasthi et al. (42) showed no relationship between the dose and the peak expansion, or exposure of CAR T-cells. However, there is discrepancy in the observation that the patients with higher pre-infusion tumor burden showed increased CAR T expansion (46), whereas Awasthi et al observed that there was no association between the baseline tumor burden and the CAR T-cell kinetics (42). The quantification of baseline tumor burden, and the location of tumor is different in these studies, so direct comparisons to or generalization of these results may not be feasible. In the same study Awasthi et al., also showed that the geometric mean of tisa-cel expansion in the peripheral blood is nearly 6-fold lower in DLBCL compared to B cell acute lymphoblastic leukemia (ALL) based on data from ELIANA clinical study. These findings may be due to CAR T-cells possibly trafficking to the loci of target cells, where the primary locations of tumor cells are in lymph nodes or the extranodal/extramedullary sites in DLBCL patients and the target accessibility, accessible antigen amount may also play a role in the differential CAR T-cell expansion. Also, the transgene levels measured in the blood in DLBCL patients may not actually represent the overall CAR T-cell expansion and may not fully reflect the CAR T-cell interaction with the target at the tumor site.

A possible solution to establish relationships between CAR T cellular kinetics, its safety and efficacy, is studying bone marrow, where $CD19^+$ B cells are continually produced, which may be more important than studying blood. Mueller et al. (46) observed a transient decline in circulating CAR T-cells immediately after peak infusion levels due to the distribution of the cells throughout the peripheral blood (PB), bone marrow (BM), and other tissues in ALL and CLL patients, which is also true in DLBCL patients (42, 48). Cellular kinetic analyses by Mueller et al. (46) showed that in pediatric B-ALL, adult ALL and CLL patients, CAR T-cells were present at higher levels in BM, for a longer time in responding patients compared to nonresponding patients. The blood to bone marrow partitioning showed that CAR T distribution in bone marrow was 44%, 67%, and 68.8% of that present in blood on day 28, and months three and six, respectively (47). These observations iterate the importance of studying the CAR T kinetics in BM and other tissues in providing more insights into factors affecting their proliferation and persistence.

CRS is an on-target toxicity and several studies showed that sCRS is associated with higher tumor burden (18, 45, 46, 64) suggesting high CAR T-cell proliferation *in vivo*; however, inconsistent results found from correlation analysis of tumor burden, CAR T kinetics and patient responses impede a thorough understanding of the underlying factors driving CAR T efficacy and safety. Locke et al. (76), using logistic regression analysis, evaluated the relationship between covariates and patient responses in ZUMA-1 DLBCL patients and found that there was a lower durable response rate in patients with higher tumor burden compared with patients with lower tumor burden, though they have comparable peak CAR T-cell levels. Whereas, Liu et al. found no association of tumor burden with patient responses in MM and CLL patients (48). Locke et al. (76) also showed that the baseline tumor burden was not associated with *in vivo* $C_{max}$ in DLBCL patients though tumor burden is associated with patient response. Whereas other studies (41, 44) have found correlation between tumor burden and increased expansion CAR T-cells. Though the predictions in Hardiansyah et al. (44) were made using a model describing the kinetics of CAR T-cells that resembles ecological predator-prey interactions, the choice of this modeling framework was driven by the kinetic data. The model is limited to the interactions between CAR T-cell and tumor cells and mechanisms considered for each cell population. Based on the "predator-prey" assumption, a relationship between tumor burden and CAR T-cells is already set and therefore, the outcomes expected are to know, for example, given certain dose of CAR T with certain amount of tumor burden, how will their interaction affect the outcome of one or the other species; or how fast (or when) will that be achieved. Thus, this model is able to describe the relationships between tumor burden and CAR T kinetics and is also validated by the observations from the clinical findings, which shows correlation between the tumor burden and CAR T kinetics.

**T-cell composition:** Immunophenotypic characterization of T-cell composition of starting leukapheresis product and the final infused T-cell product could be a tool to predict the CAR T kinetics and the response (77). Turtle et al. (61) have shown that immunotherapy with defined composition facilitated design of CART dosing that mitigated toxicity and improved disease-free survival. However, the comparisons were not made with other CAR T compositions to see how much benefit is achieved. Moreover, manufacturing defined composition of CAR T-cells is difficult, costly and challenging as most of the patients in Turtle's study (61) could not achieve the desired CAR T composition because of low lymphocyte counts. On the other hand, Mueller et al. (47) determined the CD4:CD8 ratio in the final product of B-ALL patients from ELIANA study and the relationship between clinical response, safety, and *in vivo* expansion was evaluated. Their analyses presented a similar CD4:CD8 ratio of infused CAR T between responding and non-responding patients, suggesting that in vivo expansion is independent of initial CD4:CD8 ratio.

**Conditioning therapy:** The proliferation and persistence of CAR T-cells may also be determined by the lymphodepletion or conditioning therapy that the patient received. Turtle and colleagues showed that differences in lymphodepleting regimen, for example, cyclophosphamide with or without addition of fludarabine, may result in different CAR T-cell expansion and persistence and the clinical outcome, in patients with non-Hodgkin lymphoma (61, 62). Though it is not clear how the conditioning therapy affects CAR T kinetics, reduced competition for homeostatic cytokines required for T-cell maintenance, and reduced regulatory T-cells, making the environment more hospitable for CAR T-cells could be the possible underlying mechanism (63, 77). However, Mueller et al. (47) reported that the type of lymphodepletion regimen did not affect expansion and persistence of the transgene.

**Future of modeling in CAR T-cell product development and evaluation:** Since "one dose for all" approach is not suitable for CAR T-cell therapies, mathematical models can be used to identify optimally effective and safe dose by considering the drug- and patient-specific factors. Most of the current models developed are either for academic or pharmaceutical industry use, such as for exploration of dose-response and benefit-risk assessments. The use of such models in the clinical settings as a tool for personalized therapy is still in its infancy, as CAR T-cell therapy is relatively a new field and none of the models developed to date exhibit strong predictive power and reliability to be used as a guide for making treatment decisions. The reason behind this is the insufficient data and the discordance found in the literature because of patient variability, and/or drug specific factors.

With the rapid evolution of CAR T-cell therapy, we except to have more data available which helps in further progress of modeling efforts and model comprehensiveness, to address the knowledge gaps and to guide decision making. Accumulation of the following sets of CAR T-cell therapy data would certainly beneficial for model development and validation:

(i) Cellular kinetic data of CAR T-cells in blood, bone marrow and other tissues would help in understanding the contradictory observations from ALL and DLBCL patients treated with the same

CAR T product, tisa-cel, and CAR T-cell distribution dynamics and the tumor infiltrating T-cell data would aid in CAR T-cell modeling in solid tumors (19).

(ii) Longitudinal data of all relevant cytokines, along with other patient clinical characteristics such as blood pressure, body temperature and oxygen levels, would help better predict the biomarkers of sCRS (21, 27, 44).

(iii) Longitudinal immunophenotyping data of CAR T-cells would help in understanding, the impact of T-cell subsets on the outcome, the optimum CAR T composition required for better efficacy and safety, and the reasons for conflicting observations regarding their impact (47, 61, 62).

(iv) Long term data of therapeutic outcomes of individual patients would help build survival models for better long-term predictions (8, 50).

(v) Individual patient data of immunomodulatory drugs received along with their dosing schedules, would help in better studying the impact of these drugs on CAR T-cell kinetics and in optimizing the timing of immunomodulatory drug infusion (48, 71-73).

(vi) Individual patient lymphocyte counts after lymphodepletion therapy, would help in understanding the impact of lymphodepletion on the therapeutic outcome, the underlying mechanism, and the optimal time to treat the patient with lymphodepletion therapy before CAR T-cell infusion (47, 61, 62).

(vii) To get better understanding of the impact of initial tumor burden on CAR T-cell kinetics, safety and the therapeutic outcome, tumor burden in blood and bone marrow just before CAR T-cell infusion should be reported and the quantification methods used for reported initial tumor burden across studies should be harmonized, for better inter study comparisons (42, 46, 48).

When a model is developed, factors like type of disease, CAR design and patient characteristics need to be considered as each of these factors have different intrinsic properties and the predictions made by the model should be able to validate similar independent prospective studies. Also, when building a model, the choice between a simple quantifiable model, or a detailed QSP model should be considered carefully, as the predictive power of the model can be reduced because of oversimplification or overfitting respectively (78). Hence the optimal level of the model complexity should be determined based on the purpose and the predictive power essential to capture the complex immune interactions and provide reliable estimations. Current data and the information available is too immature to make any comparisons and to generalize the findings to all types of CAR T-cell products (75). Though its challenging, achieving harmonization between clinical trials of different CAR T products would increase the value of clinical data collected and provide high quality evidence, facilitating easy decision making and validating the models.

Though, various types of models are being developed to understand the underlying mechanisms and the factors responsible for patient variability in terms of therapy outcomes, our journey towards developing personalized, safe and effective CAR T-cell therapies with the aid of modeling approach is still at the early stages. Nonetheless, such efforts certainly show potential to accelerate the development of CAR T-cell therapy for various indications in the future and these efforts can be successful with a close collaboration between the clinical researchers and modelers.

## ACKNOWLEDGEMENTS

The authors thank Gwendolyn Halford, FDA Library, for curating the literature search queries. We would like to thank Joanne Berger and Daniel Sloper, FDA Library, for editing the manuscript. This project was supported in part by an appointment to the Research Participation Program at OBE/CBER, U.S. Food and Drug Administration, administered by the Oak Ridge Institute for Science and Education through an interagency agreement between the U.S. Department of Energy and the FDA. The authors received no financial support from any source, and there is no conflict of interest.

## DISCLAIMER

This article reflects the views of the authors and should not be construed to represent the FDA's views or policies. These comments do not bind or obligate FDA.

## AUTHOR CONTRIBUTIONS

**CONCEPTUALIZATION:** UN, MRM, ONY, HY
**DATA CURATION:** UN, ONY
**FORMAL ANALYSIS:** UN, MRM, ONY, XW
**FUNDING ACQUISITION:** HY
**INVESTIGATION:** UN, MRM, ONY, XW
**METHODOLOGY:** UN, ONY
**PROJECT ADMINISTRATION:** ONY, HY
**RESOURCES:** HY
**SUPERVISION:** HY
**VISUALIZATION:** UN, MRM
**WRITING – ORIGINAL DRAFT:** UN, MRM, ONY, XW, HY
**WRITING – REVIEW & EDITING:** UN, MRM, ONY, XW, HY